\documentclass[conference]{IEEEtran}
\IEEEoverridecommandlockouts
\usepackage{cite}
\usepackage{amsmath,amssymb,amsfonts}
\usepackage{multirow}
\usepackage{subcaption}
\usepackage{url}
\usepackage{booktabs}
\usepackage{colortbl}
\usepackage[table]{xcolor}
\makeatletter
\newif\if@restonecol
\makeatother

\usepackage[linesnumbered,ruled,vlined]{algorithm2e}
\usepackage{algpseudocode}
\usepackage{graphicx}
\usepackage{textcomp}
\usepackage{xcolor}
\def\BibTeX{{\rm B\kern-.05em{\sc i\kern-.025em b}\kern-.08em
    T\kern-.1667em\lower.7ex\hbox{E}\kern-.125emX}}
\begin{document}

\title{
Stabilizing and Optimizing Inter-Shell Routing in LEO Networks with Integrated Routing Cost
}
\author{
Yaojia Wang, Qi Zhang, Kun Qiu, \textit{Senior Member, IEEE}, Yue Gao, \textit{Fellow, IEEE}
}

\maketitle

\begin{abstract}
The low Earth orbit (LEO) mega-constellation network (LMCN), which uses thousands of satellites across multi-shell architectures to deliver different services, is facing challenges in inter-shell routing stability due to dynamic network topologies and frequent inter-satellite link (ISL) switching. Existing strategies, such as the Minimum Hop Path set, prioritize minimizing hop counts to reduce latency, but ignore ISL switching costs, which leads to high instability. To overcome this, the Adaptive Path Routing Scheme introduces path similarity thresholds to reduce the ISL switching frequency between shells. However, the greedy approach of Adaptive Path Routing Scheme is often trapped in local optima, sacrificing inter-shell path distance efficiency. To address these limitations, we propose the Dynamic Programming-based Integrated Routing Cost (DP-IRC) algorithm, which is designed explicitly for inter-shell routing optimization. By formulating multi-shell paths as a multistage decision problem, DP-IRC balances hop counts and ISL stability through an Integrated Routing Cost (IRC) metric, combining inter-/intra-shell hops and switching costs. Experiments over 60 time slots with real-world Starlink and OneWeb configurations show that DP-IRC reduces inter-shell ISL switching rates by 39.1\% and 22.0\% compared to the Minimum Hop Path set strategy and Adaptive Path Routing Scheme, respectively, while still maintaining near-optimal end-to-end distances. 

\end{abstract}


\section{Introduction}
The rapid development of the low Earth orbit (LEO) mega-constellation network (LMCN) significantly contributes to several aspects of human scientific progress, such as communication, navigation, and remote sensing \cite{doi:10.34133/2022/9865174}.
LMCN has the characteristics of low delay, low propagation loss, high bandwidth, and seamless coverage, which is the cornerstone of the integrated space-air-ground network \cite{sentence1}.
In recent years, series of low orbit satellite constellations projects such as Starlink \cite{starlink} and OneWeb \cite{oneweb_coverage_tarcker} have been proposed and are gradually being deployed.
These large constellation projects apply more than thousands of satellites and provide services such as the cloud service platform and the analysis and prediction of image information from earth observation \cite{app11209490}.


Large-scale satellite constellations, such as Starlink's 42,000-satellite plan and OneWeb's 720-satellite constellation, are transforming the space environment and posing challenges on routing algorithm design, network topology dynamics, radio channel coordination, and resource management 
[\citenum{Space-Track}, \citenum{ekici2001distributed}, \citenum{wang2007topological}, \citenum{al2022next}]. 
Current research focuses on the use of LMCNs for applications such as high-precision global positioning, using advanced sensors and orbit data to achieve centimeter or millimeter accuracy \cite{aerospace12020102}. These studies aim to improve network performance and service quality by addressing challenges such as frequent ISL switching and dynamic topologies through optimized routing algorithms and resource management strategies.

In the inter-shell routing design of LMCNs, the highly dynamic network and frequent ground-satellite link (GSL) switching lead to constant ISL changes.
For example, satellite switching occurs on average every 10 minutes, whereas beam switching occurs on average every $1-2$ minutes \cite{li2005research}.
Frequent ISL switching leads to significant loss of network capacity and 8\% increased delay cost on average, while also increasing the risk of link interruptions, which negatively impacts the overall performance and reliability of satellite communication systems \cite{zhou2023link}.

The Minimum Hop Path set strategy minimizes the number of hops between two satellites, thus reducing latency \cite{wu2024enhancing}.
On the other hand, the Adaptive Path Routing Scheme focuses on decreasing the ISL switching frequency. It uses satellites' predictable movements to optimize path selection, enhancing network stability and efficiency.
Compared to traditional methods, Adaptive Path Routing Scheme decreases the ISL switching rate and delay jitter by approximately 75\% and 50\% on average, respectively \cite{Chen2003AnAP}.

Despite the advantages of Adaptive Path Routing Scheme, certain issues remain to be addressed.
First, it focuses on immediate QoS needs and current network conditions, often neglecting the broader network context and future implications, leading to suboptimal decisions.  
Second, compared to Minimum Hop Path set strategy, it prioritizes reducing ISL handoffs by selecting longer paths, but this increases transmission delays and prevents truly optimal path selection. Although Adaptive Path Routing Scheme achieves fewer handoffs, the trade-off compromises overall efficiency\cite{Chen2003AnAP}.


To address the limitations of existing algorithms, we propose the Dynamic Programming-based Integrated Routing Cost (DP-IRC) algorithm. DP-IRC supports multi-access terminals and introduces path diversity for enhanced reliability. By modeling routing as a multi-stage decision problem and applying dynamic programming, DP-IRC optimizes paths using an Integrated Routing Cost (IRC) metric. The IRC combines hop count for transmission efficiency and ISL switching costs for topology stability, weighted by \(\alpha\) and \(\beta\) (\(\alpha + \beta = 1\)). This approach ensures globally optimal routing decisions, avoids local optima of greedy methods, reduces ISL switching rates, and prevents excessive path lengths. The experimental results show that DP-IRC reduces ISL switching rates by 39.1\% and 22.0\% compared to the Minimum Hop Path strategy and the Adaptive Path Routing Scheme, respectively.
Briefly, this paper makes the following contributions:
\begin{itemize}
    \item 
    We deployed the existing inter-satellite routing design to inter-shell routing design and studied the problem of frequent ISL switching in it.
    \item We propose the IRC metric which quantifies the trade-off between path efficiency and stability and is formalized through rigorous mathematical analysis.
    \item Experimental validation using real-world Starlink and OneWeb configurations, demonstrating the superiority of DP-IRC in ISL switching rate reduction, load balancing, and scalability.
\end{itemize}

The rest of the paper is organized as follows: Section II introduces the background and our motivations. Section III presents the overview design, and our algorithm is explained in Section IV. Then, Section V describes our experiments and analyses the results. Finally, Section VI concludes the whole work.

\section{Background}





\subsection{Constellation Architecture and Link Dynamics}

Modern low Earth orbit (LEO) mega-constellations adopt multi-shell architectures to achieve global coverage and service redundancy. Each shell consists of satellites that operate at a specific altitude and inclination. For instance, Starlink's \textit{shell 1} comprises 1,584 satellites at 550 km altitude with 53° inclination, while OneWeb's \textit{primary shell} contains 720 satellites at 1,200 km altitude with 87.9° inclination \cite{starlink}, \cite{oneweb_coverage_tarcker}. 

\begin{figure}[]
\centerline{\includegraphics[width=0.6\linewidth]{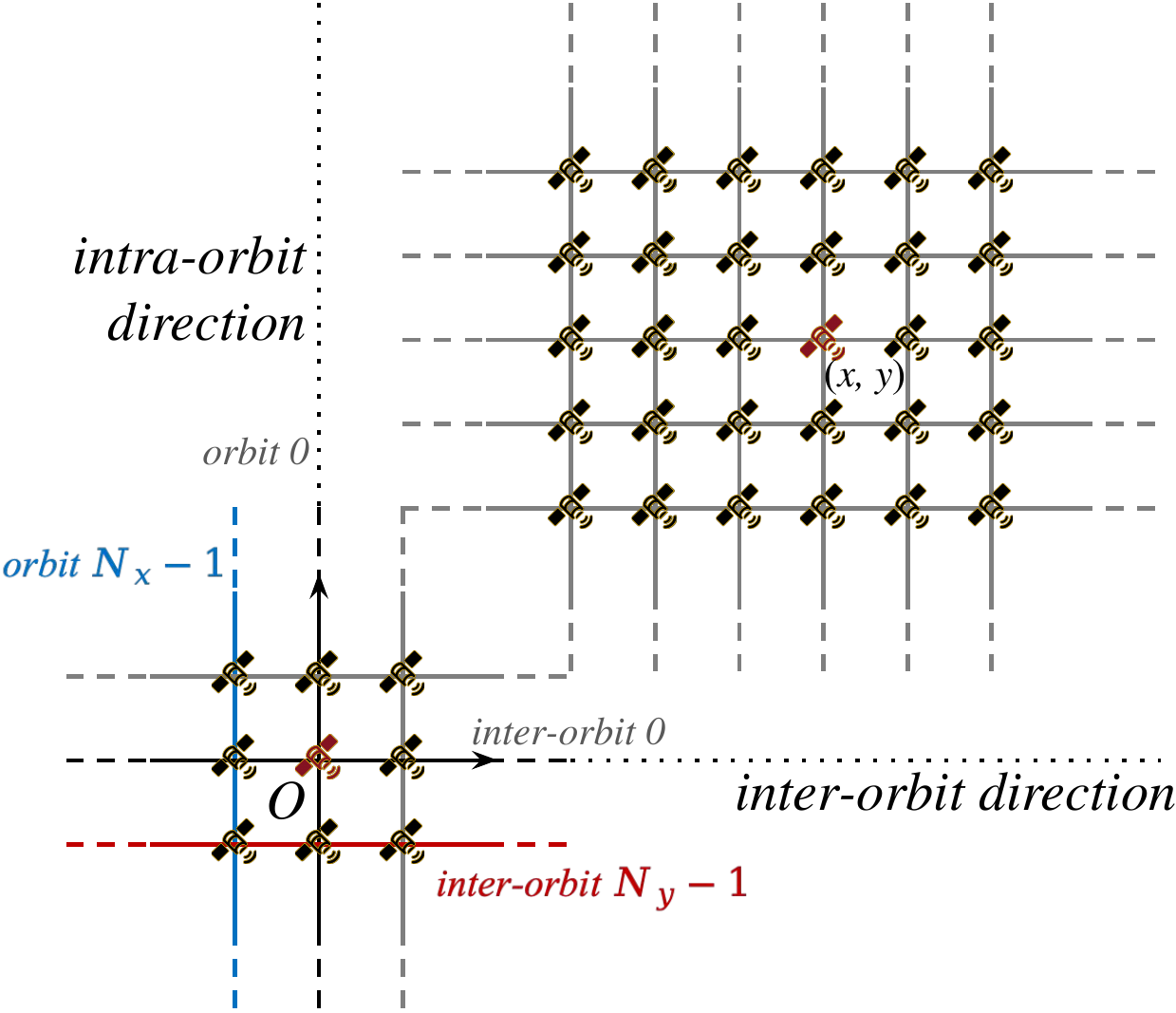}}
\caption{The satellite numbering system is placed in a coordinate system based on the +Grid inter-satellite topology.}
\label{fig:coordinatesystem}
\end{figure}

For ISL-based satellite communication, the most studied topology is the +Grid topology in which each satellite has links to the two nearest satellites in its own orbit and to the two nearest satellites in adjacent orbits on either side, as shown in Fig. \ref{fig:coordinatesystem} \cite{zhang2024depth}. This structured connectivity enables critical deterministic hop count calculations for routing optimization.

In current configurations, a ground station (GS) typically connects to one satellite per shell. This connection dynamically changes as the satellites move, ensuring continuous communication and supporting data transmission between the shells \cite{yoon2023navigating}.


When calculating inter-satellite routing, we focus on hop count rather than actual distance. Research shows that nearly all Shortest Distance Paths (SDPs) fall within the Minimum Hop Path set, meaning that SDPs can be simplified to the Minimum Hop Path region without significant loss of optimality \cite{SDPandMHP}. This approach reduces computational complexity while maintaining routing efficiency and network performance.

\subsection{Hop Calculation and ISL Switching Rate}





In the +Grid topology, satellites are mapped to a 2D grid using their global index \( S \): \( (x, y) = (\lfloor S/N_y \rfloor, S \mod N_y) \), where \( x \in [0, N_x-1] \) (orbital plane) and \( y \in [0, N_y-1] \) (in-orbit position). For two satellites with indices \( S_1, S_2 \), the minimal hop count between them is:  
\[
\text{Total Hops} = \min(|\Delta x|, N_x - |\Delta x|) + \min(|\Delta y|, N_y - |\Delta y|)
\]  

where \( \Delta x = \lfloor S_2/N_y \rfloor - \lfloor S_1/N_y \rfloor \), \( \Delta y = (S_2 \mod N_y) - (S_1 \mod N_y) \).

ISL switching cost \( \Delta_{\text{ISL}} \) (total hop changes between consecutive time slots \( t \) and \( t-1 \)) is:  
\[
\Delta_{\text{ISL}}(P_t, P_{t-1}) = \left| \text{Hops}_x^t - \text{Hops}_x^{t-1} \right| + \left| \text{Hops}_y^t - \text{Hops}_y^{t-1} \right|
\]  

where \( \text{Hops}_x^t = \min(|\Delta x^t|, N_x - |\Delta x^t|) \), \( \text{Hops}_y^t = \min(|\Delta y^t|, N_y - |\Delta y^t|) \), and superscripts denote time slots.

ISL switching rate is the ratio of \( \Delta_{\text{ISL}} \) to the previous total hops:  
\[
\text{ISL switching rate} = \frac{\Delta_{\text{ISL}}(P_t, P_{t-1})}{\text{Total Hops}_{t-1}}
\]


This metric quantifies the severity of changes in routing paths over time, with a higher value indicating more significant changes to ISLs.




\begin{figure*}[htbp]
    \centering
    \includegraphics[width=0.75\textwidth]{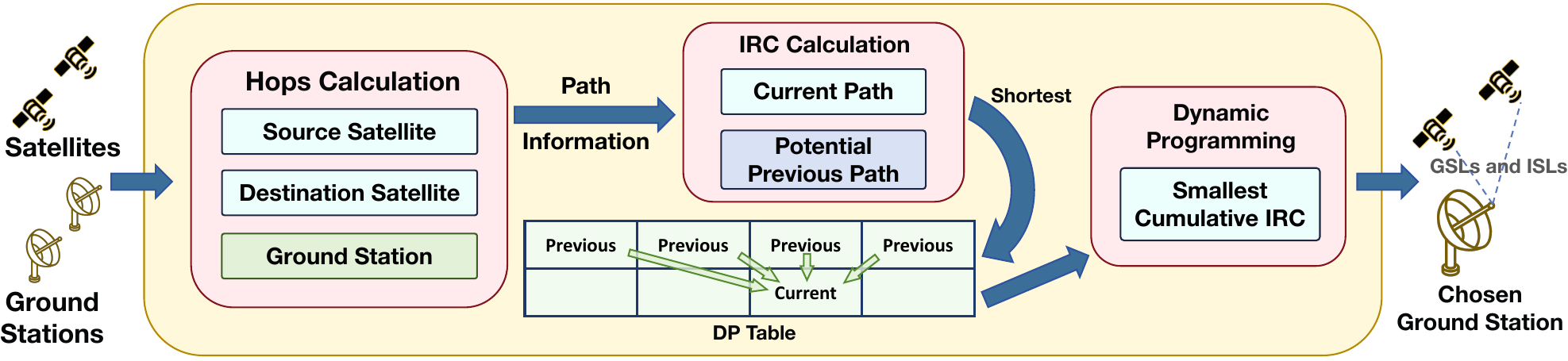} 
    \caption{Overview of DP-IRC}
    \label{fig:DPIRC_overview}
\end{figure*}

\subsection{Existing Routing Strategies and Limitations}

The following are existing routing strategies: 
\begin{itemize}
    \item Minimum Hop Path set strategy selects the path with the fewest ISL hops but ignores switching costs. Although low latency and shortest path are achieved, it generates high ISL switching rates.
    \item  Adaptive Path Routing Scheme reduces switching by enforcing specific path similarity between consecutive slots\cite{Chen2003AnAP}. However, its greedy approach is often trapped in local optima, sacrificing path efficiency
    or stability when similarity thresholds fail.
\end{itemize}

Minimum Hop Path set strategy represents a fundamental approach (minimizing the number of hops) to routing in satellite networks.
This simplicity makes it a useful reference point for understanding basic performance metrics, such as latency. Therefore, we choose the Minimum Hop Path set strategy as the baseline method for comparison.

We select the Adaptive Path Routing Scheme for comparison as it represents a strategy prioritizing ISL switching frequency reduction, crucial for LEO constellations' stability and resource use. However, its greedy heuristic nature can cause suboptimal solutions, longer cumulative distances, and instability when similarity thresholds are not met.

To comprehensively assess these strategies, we simulate through 60 time slots, capturing long-term algorithm performance, and revealing how each strategy balances latency, switching costs, and stability in dynamic LEO environments.








\section{Overall Design}
The overall design of DP-IRC is shown in Fig. \ref{fig:DPIRC_overview} 
The primary goal is to find the optimal path for data transmission between two satellites from two shells, considering the dynamic nature of the network and the constraints imposed by the satellite orbits and the locations of the ground stations. 
DP-IRC has several core components, namely, \textit{Inter-satellite hops calculation}, \textit{Integrated routing cost calculation}, and \textit{Dynamic Programming for Path Optimization}.


\subsection{Inter-satellite Hops Calculation}
In order to calculate the smallest hop count between the two satellites when each ground station is chosen, we iterate through the ground stations and calculate the shortest paths between two satellites, considering the connections through ground stations. In this part, we compute the distances and path differences for each potential route and store the data for the next two periods.

\subsection{Integrated Routing Cost Calculation}
To comprehensively evaluate path quality, we propose the Integrated Routing Cost (IRC) parameter, which quantifies both transmission efficiency and network stability. The IRC value combines the current path's hop count with the switching cost from previous connections, where lower values indicate superior paths. The IRC calculation method will be used in the dynamic programming algorithm to obtain IRC values between all candidate paths and their former paths at each time step.

\subsection{Dynamic Programming for Path Optimization}
The dynamic programming approach is used to optimize the path selection process. In this part, we construct a cost matrix and use it to determine the optimal path from the starting satellite to the destination satellite over a series of time steps, enabling global optimization through backward
cost propagation.

\vspace{-10pt}
\section{Algorithm Design}

\subsection{Symbols and Definitions}
The key symbols used in this paper and their definitions are shown in TABLE \ref{tab1}. 

\begin{table}[]
\vspace{15pt}
\caption{SYMBOLS OF DEFINITIONS}
\begin{center}
\begin{tabular}{|c|l|}
\hline
Notation&Definition \\
\hline
 $dp$&Dynamic programming table\\
 $N_x^h$&Number of orbital planes in $h$\\
 $N_y^h$&Satellites per plane in $h$\\
 \hline
 $h$& Satellite shell\\
 $T$&Number of time slots\\
 $t_j$&Time slot $j$\\
 $S_h$&Set of satellites in $h$\\
 $s_{i}^h$&Satellite $i$ in $h$\\
 $G$&Set of ground stations\\
 $g_i$&Ground station $i$\\
 $L$&Set of GSLs\\
 $l_{i}^h$&GSLs between $g_i$ and $h$\\
 $e_i^h$&Satellites linked to $g_i$ in $h$\\
 $D$ &List of routing data\\
 $d_j$&Routing data at $t_j$\\
 $p_i^j$&Routing data for $g_i$ at $t_j$\\
 $\alpha$&Weight of hop count of path\\
 $\beta$&Weight of ISL variation\\
 \hline
\end{tabular}
\label{tab1}
\end{center}
\vspace{-18pt}
\end{table}
$h$ denotes the satellite shell (A or B), and $T$ is a set of evenly divided time slots.  
$S_h$ and $G$ represent satellites in shell $h$ and ground stations, respectively.  
$L_h$ includes GSLs (Ground-Satellite Links) connecting the ground station $g_i$ to the satellite $e_i^h$ in shell $h$.  
$D$ stores time-slot-specific routing data $d_j$, detailing inter/intra-shell paths between satellites.  
$p_i^j$ records $g_i$’s routing metrics (hop counts) at time slot $t_j$.  
The IRC metric balances the path efficiency (weight $\alpha$) and topology stability (weight $\beta$), with $\alpha + \beta = 1$.

\subsection{Theoretical Analysis}
In order to support our algorithm, we have the following proposition.

\textit{\textbf{Proposition}:} 
The path quality can be evaluated by the Integrated Routing Cost (IRC) parameter, which is the weighted combination of hop count and switching cost from previous connections.

\textit{\textbf{Proof}:}  
To holistically evaluate routing quality, we define the IRC metric for path $P_t$ as:  
\begin{equation}
\text{IRC}(P_t, P_{t-1}) = \alpha \cdot \underbrace{H(P_t)}_{\text{Current Path Hops}} + \beta \cdot \underbrace{\Delta_{\text{ISL}}(P_t, P_{t-1})}_{\text{ISL Switching Cost}}
\label{eq9}
\end{equation}  

where $H(P_t)$ indicates the total hop count of path $P_t$, $\Delta_{\text{ISL}}(P_t, P_{t-1})$ indicates the ISL variation between two paths.
$\alpha, \beta$ are tunable weights that balance efficiency and stability. 
If the application prioritizes transmission efficiency (e.g., real-time communication services requiring low latency), $\alpha$ should be set higher. If the application focuses more on routing stability (e.g., services sensitive to link interruptions), $\beta$ should be increased. IRC represents the comprehensive cost incurred when selecting a particular path, as it integrates both the efficiency-related cost and the stability-related cost into a single metric.

This proposition will support our algorithm description and experiment in Section V, proving the optimality of the results and the rationality of the classification operation.

\subsection{Algorithm Description}
\subsubsection{DP-IRC}

DP-IRC algorithm consists of three parts: path calculation, IRC calculation for the optimal path, and dynamic programming.
\begin{figure}
    \centering
    \includegraphics[width=0.7\linewidth]{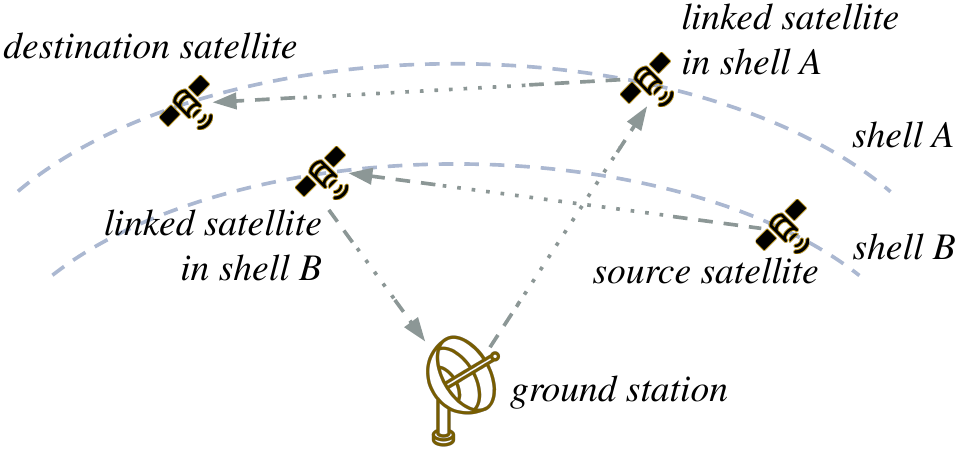}
    \caption{Signals between source and destination satellites are routed via ISL hops to their respective same-shell ground station-linked satellites, enabling ground stations to relay cross-shell signals through dual-shell satellite connections.}
    \label{fig:inter-shell}
\end{figure}

\begin{algorithm}
  \caption{Path Calculation}
    \label{alg:initialization}
  \KwIn{$s_m^A$, $s_n^B$, $N_x^A$, $N_y^B$, $N_x^B$, $N_y^B$, $L$, $G$}
  \KwOut{$d_i$}
  $d_j = \emptyset$\\
  \For{$i$ from $0$ to $length(G)-1$}
  {
    calculate $\Delta x_A$ and $\Delta y_A$ as inter-orbit hops and intra-orbit hops between $s_m^A$ and $e_i^A$, respectively.\\
    calculate $\Delta x_B$ and $\Delta y_B$ as inter-orbit hops and intra-orbit hops between $s_n^B$ and $e_i^B$, respectively.\\
    $dist = \Delta x_A + \Delta y_A +\Delta x_B + \Delta y_B$\\
    $d_j = d_j \cup \{\{\Delta x_A, \Delta y_A, \Delta x_B, \Delta y_B, dist\}\}$
  }
  return $d_i$
\end{algorithm}

The path calculation Algorithm \ref{alg:initialization} is to determine the shortest path between two satellites from different shells through each ground station. The routing path is shown in Fig. \ref{fig:inter-shell}.

The dynamic programming algorithm (Algorithm \ref{alg:core}) computes the optimal path for each time slot \( t_j \) using the IRC metric. This algorithm initializes a DP table \( dp \) and a previous index table \( prev \),
sets the first row of $dp$ based on initial distances and iteratively fills the table by calculating transition costs and updating the minimum costs. After determining the minimum cost in the final row, it retraces through $prev$ to construct the optimal path.
\begin{algorithm}

  \caption{Dynamic Programming}
  \label{alg:core}

  \KwIn{$D$, $G$}
  \KwOut{$path$}

         $T$ = length($D$)\\                
         $N$ = length($G$)\\             
         $dp$ = $T \times N$ matrix filled with $\infty$\\  
         $prev$ = $T \times N$ matrix filled with $-1$\\  

     \For{$p_i^0\in d_0$}
        {
        let $dp[0][i]$ be the $(dist \ \text{in}\ p_i^0) \times \alpha$\\
        }       
    \For{$d_j\in D-\{d_0\}$}
    {
        \For{$p_i^j \in d_j$}
        {
            \For{$p_k^{j-1} \in d_{j-1} $}
            {
                calculate $c$ as the $\text{IRC}(p_i^j,p_k^{j-1})$ between $p_i^j$ and $p_k^{j-1}$\\
                \If{$dp[j-1][k]+c<dp[j][i]$}
                {
                    $dp[j][i] = dp[j-1][k]+c$\\
                    $prev[j][i] = k$
                }
            }
        }
    }

    $path = \emptyset$\\

    $m =$ the index minimum value in $dp[T-1]$\\

    \For{$j$ from $T-1$ to $0$}
    {
        $path = path \cup p_m^j$\\
        $m = prev[j][m]$
    }

    reverse $path$\\
    return $path$
\end{algorithm}



\subsubsection{Proof of Optimality}
We prove that the DP-IRC algorithm finds the globally optimal path by minimizing the cumulative Integrated Routing Cost (IRC) across all time slots, leveraging dynamic programming properties.  
The DP table is initialized with the IRC values of the initial paths, calculated as \( \alpha \cdot H(P_{t_0}) \) (no prior path, so \( \beta \cdot \Delta_{\text{ISL}} = 0 \)). The minimum value here corresponds to the optimal path for \( t_0 \) by the definition of IRC.  
Assume optimality up to \( t_{k-1} \). For \( t_k \), the algorithm evaluates all transitions from the \( t_{k-1} \) paths, calculating the total IRC as \( \text{cumulative IRC}_{t_{k-1}} + \alpha \cdot H(P_{t_k}) + \beta \cdot \Delta_{\text{ISL}}(P_{t_k}, P_{t_{k-1}}) \). Updating the DP table with the minimum transition cost preserves optimality for \( t_k \).  
By induction, the algorithm minimizes cumulative IRC in all slots. The backtracking of \( t_{T-1} \) yields the globally optimal path, confirming the optimality in balancing hop count and switching cost.

\subsection{Example Analysis}
\begin{figure}[]
    \centering
    \includegraphics[width=0.6\linewidth]{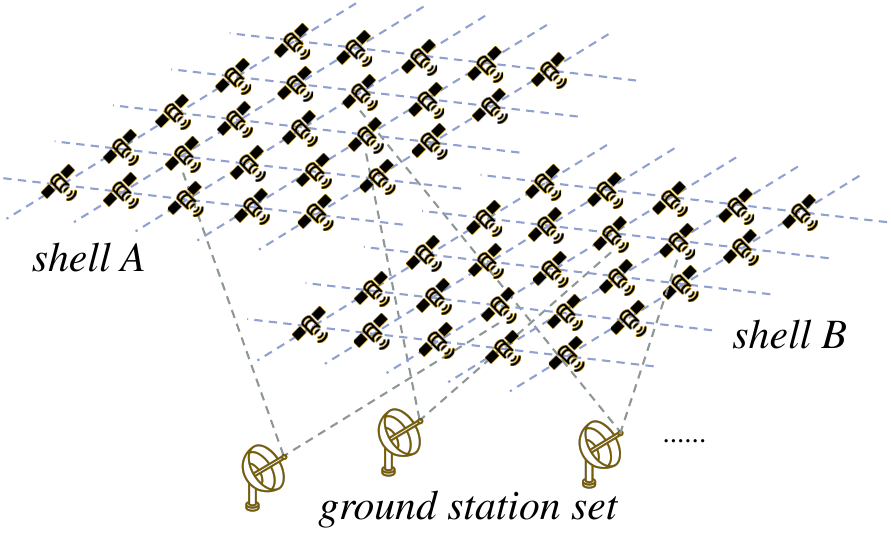}
    \caption{Initially, every ground station connect one satellite to each of the two shells.}
    \label{fig:enter-label}
\end{figure}


As shown in Fig. \ref{fig:enter-label}, our set-up includes two satellite shells (A and B) and a set of ground stations G. Each ground station connects to one satellite in each shell. We choose a source satellite from shell A and a destination satellite from shell B. Based on GSL data over time, we first initialize by computing the paths between the two satellites through each ground station. Then, the dynamic programming algorithm determines the optimal path for each connection scenario across all time slots.

\begin{figure}
    \centering
    \begin{subfigure}{0.24\textwidth}
        \centering
        \includegraphics[width=\linewidth]{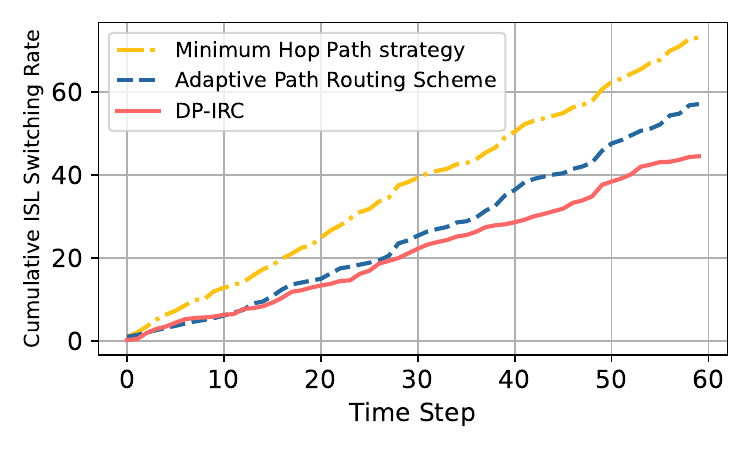}
        \caption{ISL switching rate}
        \label{fig:ISL_switching rate}
    \end{subfigure}
    \hfill
    \begin{subfigure}{0.24\textwidth}
        \centering
        \includegraphics[width=\linewidth]{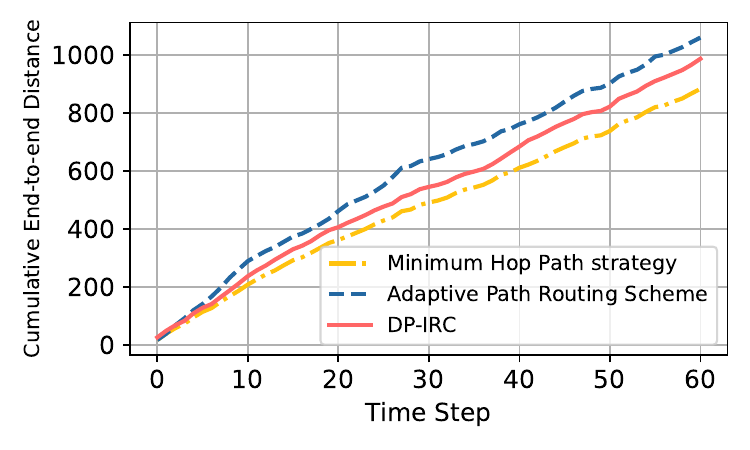}
        \caption{End-to-end distance}
        \label{fig:path distance}
    \end{subfigure}
    \vspace{15pt}
    \caption{Cumulative ISL switching rate and End-to-end distance}
    \vspace{5pt}
\end{figure}

\section{Experiments}
\subsection{Experiment Setup}

In our experiments, we chose Starlink \textit{shell 1} \cite{starlink} and OneWeb \textit{primary shell} \cite{oneweb_coverage_tarcker} for the study. In addition, we selected 165 ground stations registered by Starlink around the world as ground nodes \cite{starlink}. To ensure the adaptability of the algorithm to real satellite constellation scenarios, we conducted experiments using the actual structure of the constellation derived from TLE data published by Starlink \cite{starlink} and OneWeb \cite{liu2022starlink}. TABLE \ref{tab:shell_parameter} shows the detailed configurations.
\\
\vspace{2pt}
\begin{table}[h]
    \renewcommand{\arraystretch}{1.3} 
    \caption{SHELL PARAMETER CONFIGURATION}
    \centering
     \begin{tabular}{|c|c|c|}
    \hline
     Shell&Parameters&Value \\ \hline
    \multirow{2}{*}{Starlink shell 1}&Orbit number ($N_x^A$) &72\\  
    \cline{2-3}  
     &In-orbit satellite number ($N_y^A$)&22\\ \hline
    \multirow{2}{*}{OneWeb primary shell} & Orbit number ($N_x^B$)&18 \\  
    \cline{2-3}  
    & In-orbit satellite number ($N_y^B$)& 40\\ \hline
    \end{tabular}
        \label{tab:shell_parameter}
\end{table}
\vspace{-3pt}

Our experiments focus on the Minimum Hop Path set strategy that only takes into account the length of the path, the Adaptive Path Routing Scheme which takes ISL switches into account and computes the optimal path in a greedy way\cite{Chen2003AnAP}, and our proposed DP-IRC algorithm.
The above algorithms are compared and analyzed mainly on the basis of path stability, ground station load balance, and cumulative end-to-end distance.

We employ the C-LRST algorithm\cite{lwh} to determine the GSL configuration between the ground station set and Starlink \textit{shell 1}, OneWeb primary shell, with a 5-minute interval between the GSL switches. The results of the experiment are shown in TABLE \ref{tab:result}.

\begin{table*}[]
\centering
\renewcommand{\arraystretch}{1.2} 
\caption{RESULTS OF THE EXPERIMENT}
\label{tab:result}
\begin{tabular}{|c|c|c|c|c|c|c|}
\hline
\textbf{Constellation Scale}& \multicolumn{1}{c|}{\textbf{large scale}} & \multicolumn{5}{c|}{\textbf{normal scale}} \\ \hline
 \textbf{Experimental Index} (Average Value)& \multicolumn{2}{c|}{\textbf{ISL switching rate} }& \textbf{end-to-end distance}& \multicolumn{3}{c|}{\textbf{Ground Station load variance}}\\ \hline
 \textbf{Ground Station Number}& \multicolumn{4}{c|}{\textbf{165}}& \textbf{110}& \textbf{55}\\ \hline
\textbf{DP-IRC} & 0.78771 & 0.74249 & 16.19672& 0.50& 0.85 & 1.66 \\ \hline
\textbf{Minimum Hop Path set strategy} & /& 1.21962 & 14.49180 & 0.49& 0.79 & 1.84 \\ \hline
\textbf{Adaptive Path Routing Scheme} & /& 0.95233 & 17.39344 & 0.56& 0.88 & 2.10 \\ \hline
\end{tabular}
\vspace{-5pt}
\end{table*}

\subsection{End-to-end Path Analysis}
In this section, we first focus on communication between two specific satellites from different shells to understand the performance of the three algorithms in detail and analyze the phenomenon and causes of various indicators. Next, we study how communication between satellites in different shells varies with the number of ground stations to have a more comprehensive understanding of these algorithms.

Based on the C-LRST algorithm, we configure ground station links (GSLs) between each ground station and a designated satellite in each shell, followed by a simulation of 60 time slots. The GSL configurations are dynamically updated at each time slot, with an update interval of 5 minutes to reflect real-world operational constraints.
The source and destination satellites used in the experiment are Starlink-550 1 and OneWeb-1200 159. In the experiment, $\alpha$ and $\beta$ are set to 0.5. This setting aims to simply balance efficiency and stability, making the algorithm perform in scenarios where neither low latency nor high stability is prioritized, thus demonstrating the trade-off capability of DP-IRC.

\subsubsection{ISL switching rate}

\begin{figure}
    \centering
    \begin{subfigure}{0.24\textwidth}
        \centering
        \includegraphics[width=\linewidth]{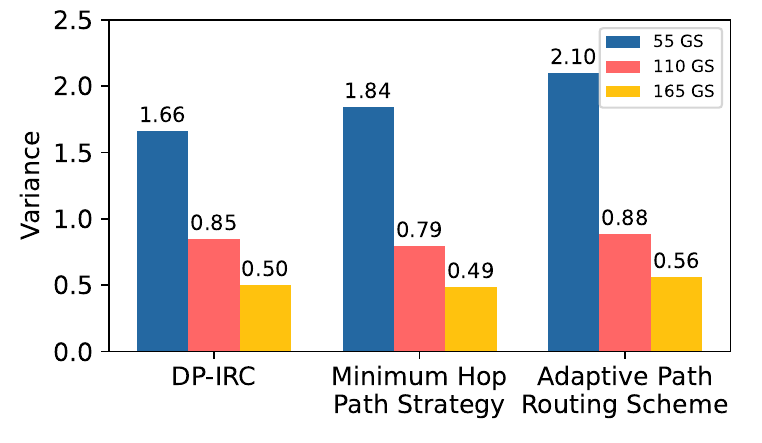}
        
        \caption{Ground station load variance}
        \label{fig:gs utilization}
    \end{subfigure}
    \hfill
    \begin{subfigure}{0.24\textwidth}
        \centering
        \includegraphics[width=\linewidth]{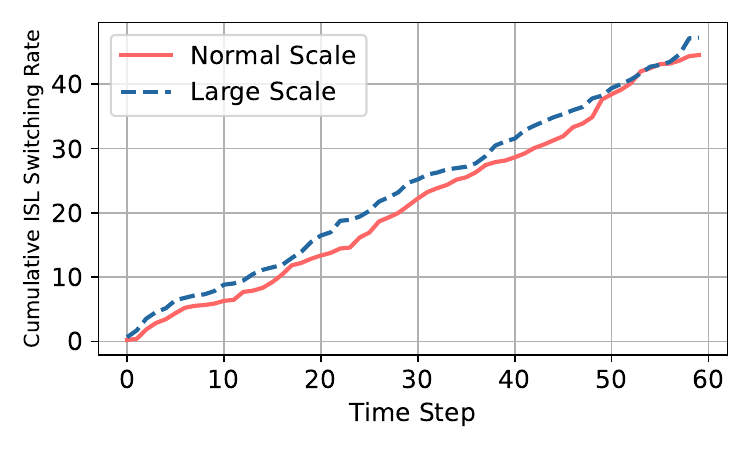}
        \caption{Large scale ISL switching rate}
        \label{fig:scalbility}
    \end{subfigure}
    \vspace{15pt}
    \caption{GS load variance and ISL switching rate at large scale}
    \vspace{5pt}
\end{figure}

The ISL switching rate, a critical metric for routing stability, reflects how frequent path changes degrade latency and reliability. As shown in Fig. \ref{fig:ISL_switching rate}, DP-IRC achieves the lowest average ISL switching rate (0.742), outperforming Adaptive Path Routing Scheme (0.952) and Minimum Hop Path set strategy (1.220). This is attributed to the global optimization mechanism of DP-IRC, which minimizes cumulative switching costs by evaluating all potential path transitions across time slots. Through backward cost propagation, it ensures smooth path evolution, avoiding abrupt changes.

In contrast, the localized path selection of Minimum Hop Path set strategy prioritizes instant hop minimization, causing frequent fluctuations. Although Adaptive Path Routing Scheme enforces a 60\% path similarity threshold to reduce switching, it selects the shortest path when thresholds fail, incurring suboptimal global stability and higher switching costs. 


\subsubsection{Cumulative end-to-end distance}
The cumulative end-to-end distance reflects the total efficiency of data transmission. As shown in Fig. \ref{fig:path distance}, Minimum Hop Path set strategy achieves the shortest distance by prioritizing instantaneous hop minimization, but sacrifices stability. DP-IRC balances trade-offs, ranking second with cumulative distances 18.3\% lower than Adaptive Path Routing Scheme, demonstrating its ability to harmonize path efficiency and switching costs.

Adaptive Path Routing Scheme’s 60\% path similarity threshold enforces stability but forces suboptimal routing, accumulating distance penalties over time. In contrast, DP-IRC’s global optimization explores diverse paths, systematically balancing distance and switching costs without rigid constraints.

\subsubsection{Ground station load balance}

As shown in Fig. \ref{fig:gs utilization}, DP-IRC achieves a better load balance on varying scales of the ground station. In smaller networks (55 GS), it attains the lowest variance (1.66), significantly outperforming Minimum Hop Path set strategy (1.84) and Adaptive Path Routing Scheme (2.10). At full scale (165 GS), DP-IRC matches the efficiency of Minimum Hop Path set strategy while surpassing Adaptive Path Routing Scheme. These results highlight DP-IRC’s inherent capability to promote load diversity through global optimization, contrasting with the inconsistent localized optimizations of Minimum Hop Path set strategy and the rigidity of Adaptive Path Routing Scheme. Its balanced stability and scalability make it ideal for dynamic satellite networks.

\subsection{Scalability}
We assessed DP-IRC's scalability via comparative experiments with baseline and expanded constellation configurations. The expanded setup doubled the number of satellites per Starlink orbit and OneWeb orbit, increasing the total number of satellites twofold.

Figure \ref{fig:scalbility} shows that for 60 time steps, both configurations had similar ISL switching rates (0.74249 for normal scale and 0.78771 for large scale). This similarity is due to the maintained +Grid topology in the expanded constellation, which preserves stable inter-satellite connectivity despite scaling up.

These results show that DP-IRC works well in large-scale constellations. Its global optimization framework lets it handle potential increases in ISL switching rate from constellation expansion, proving its strong scalability.


\subsection{Overhead and Deployment}

The computational complexity of DP-IRC is \( O(T \times G^2) \) for \( T \) time slots and \( G \) ground stations, as it evaluates all path transitions across slots to update the DP table, higher than heuristic methods such as Minimum Hop Path set (\( O(G) \)) and Adaptive Path Routing Scheme (\( O(G) \)) due to its global optimization focus.  

Modern satellite and ground station hardware (e.g., Starlink, OneWeb) can handle this via parallel computing, distributing tasks across nodes to ensure real-time performance. 
Notably, DP-IRC’s optimality is limited to a predefined time window (e.g., 60 slots). Beyond this, dynamic orbital changes require periodic re-computation to maintain optimality. The overhead is justified by gains in stability and efficiency, suiting large LEO constellations.

\section{Conclusion}

In this paper, we propose a Dynamic Programming-based Integrated Routing Cost (DP-IRC) algorithm to optimize the trade-off between path efficiency and inter-satellite link (ISL) stability in LMCN. Using multi-time slot backward cost propagation, the algorithm achieves global optimization, reducing ISL switching rates by 22.0\% and 39.1\% compared to Adaptive Path Routing Scheme and Minimum Hop Path set strategy, respectively, while maintaining low transmission latency and balanced ground station loads. Experiments on real-world Starlink and OneWeb constellations validated its scalability and adaptability in large-scale networks, demonstrating its potential for next-generation satellite communication systems. Future work will focus on real-time implementation and integration with emerging network architectures.


\bibliography{reference}

\end{document}